\RequirePackage{snapshot}% <-- Optional, for use with bundledoc only 
\documentclass[aps,prl,twocolumn,superscriptaddress,nofootinbib,showpacs]{revtex4-1}

\usepackage[english]{babel}
\usepackage{amsmath}
\usepackage{amsfonts}
\usepackage{amssymb}
\usepackage{amsthm}
\usepackage{graphicx}
\usepackage{braket}
\usepackage{todonotes}

\begin{document}

\title{Suppression of multiphoton resonances in driven quantum systems via pulse shape optimization}
\author{Denis Gagnon}
\email{denis.gagnon@uwaterloo.ca}
\author{Fran\c{c}ois Fillion-Gourdeau}
\affiliation{Universit\'e du Qu\'ebec, INRS--\'Energie, Mat\'eriaux et T\'el\'ecommunications, Varennes, Qu\'ebec, Canada, J3X 1S2}
\affiliation{Institute for Quantum Computing, University of Waterloo, Waterloo, Ontario, Canada, N2L 3G1}
\author{Joey Dumont}
\affiliation{Universit\'e du Qu\'ebec, INRS--\'Energie, Mat\'eriaux et T\'el\'ecommunications, Varennes, Qu\'ebec, Canada, J3X 1S2}
\author{Catherine Lefebvre}
\author{Steve MacLean}
\email{steve.maclean@uwaterloo.ca}
\affiliation{Universit\'e du Qu\'ebec, INRS--\'Energie, Mat\'eriaux et T\'el\'ecommunications, Varennes, Qu\'ebec, Canada, J3X 1S2}
\affiliation{Institute for Quantum Computing, University of Waterloo, Waterloo, Ontario, Canada, N2L 3G1}
\date{\today}

\begin{abstract}
	
	This Letter demonstrates control over multiphoton absorption processes in driven two-level systems, which include for example superconducting qubits or laser-irradiated graphene, through spectral shaping of the driving pulse.
	Starting from calculations based on Floquet theory, we use differential evolution, a general purpose optimization algorithm, to find the Fourier coefficients of the driving function that suppress a given multiphoton resonance in the strong field-regime.
	We show that the suppression of the transition probability is due to the coherent superposition of high-order Fourier harmonics which closes the dynamical gap between the Floquet states of the two-level system.
	%Applications of the proposed methodology include the design of high-fidelity quantum gates and the selective control of scattering mechanisms in graphene.	
	
\end{abstract}

\maketitle

Multiphoton absorption processes can be observed in quantum systems driven by the application of a sufficiently strong time-dependent external field, such as a harmonic signal.
While atoms and molecules in laser fields \cite{Chu2004} have historically been the natural testbed for the study of multiphoton processes, the rise of strongly driven systems in the fields of quantum information \cite{Son2009, Jooya2016} and condensed-matter physics \cite{Kelardeh2015, Kelardeh2016, Fillion-Gourdeau2016} has given new impetus to this field of study.
All these systems exhibit rich physics in the multiphoton regime, a prime example being the phenomenon of coherent destruction of tunneling (CDT).
This phenomenon corresponds to destructive quantum interference, which results in the suppression of Rabi population oscillations for specific values of the driving field amplitude and frequency \cite{Grossmann1991, Grifoni1998}.
In the context of quantum information, for example, CDT corresponds to the situation where a qubit does not transition to the excited state over some range of tunnel splittings even though the driving function is non-trivial.
Another system in which CDT can be realized is laser irradiated graphene, which was studied in recent works from both a time-dependent \cite{Fillion-Gourdeau2016} and Floquet perspective \cite{Gagnon2017}.
The main interest of mono-layer graphene as a two-level system is that multiphoton transitions in the strong-field regime correspond to accessible laser frequencies (in the THz or optical range), owing to the small value of the Fermi velocity and the absence of an electronic gap.
Multiphoton experiments, however, could be more readily performed in superconducting qubits because of limits on carrier lifetime in intrinsic graphene \cite{Gierz2016, Yu2016}.

The main motivation behind this Letter is the possibility of controlling dynamical quantum systems using time-dependent fields, a topic which has wide-reaching implications in several sub-disciplines of physics \cite{Brif2010, Hildner2011, Song2016, Sinitsyn2017}.
For example, Campos \emph{et al.} have recently shown that it is possible to induce the same optical response from two different atomic systems by tailoring the incident optical pulse \cite{Campos2017}.
In the present work, the focus is specifically placed on time-dependent quantum two-level systems (TLS).
Since the TLS Hamiltonian is of the generic form describing driven double-well potentials, the conclusions of this work can be applied to several quantum systems including superconducting flux qubits \cite{Oliver2005, Son2009, Jooya2016}, semiconductor superlattices \cite{Holthaus1992, Rotvig1995, Platero2004} and atomic and molecular systems \cite{Grifoni1998, Kierig2008, Shevchenko20101} (see supplementary material).
There has been much work in recent years on ``shortcuts to adiabaticity'', a set of methods which allow for the control of the final state of a quantum system in a much shorter time than a slow, adiabatic process \cite{Berry2009, Ruschhaupt2012, Torrontegui2013, Deffner2014, Baksic2016}.
Deffner has recently described this control method in Dirac materials, which include graphene and $d$-wave superconductors \cite{Deffner2015}.
In this Letter, we adopt a slightly different approach to quantum control by considering driven TLS from the viewpoint of Floquet theory \cite{Bartels2013}.
More specifically, we seek to minimize the magnitude of a physical observable -- the Floquet transition probability -- over a large energy window while driving the system with a strong, non-trivial excitation.
This allows for the selective suppression of multiphoton resonances in the absorption spectrum of the driven system.

While the suppression of a single energy, or qubit state, can be readily achieved \cite{Peirce1988, DAlessandro2001, Wu2002, Bartels2013} using optimal control theory \cite{DAlessandro2007, Glaser2015}, the suppression of multiphoton absorption over a wide spectral range is more involved.
In fact, since the optimization objective has to be defined over a range of energy parameters of the two-level Hamiltonian, one would have to use \textit{ensemble} control theory \cite{Li2006, Li2009, Li2011, Leghtas2011, Chittaro2017}.
Ensemble control refers to a set of well-established methods for the control of a continuum of quantum systems using a single applied field \cite{Glaser2015}.
In this work, we use an alternate approach from the Floquet point-of-view and resort to a high-level optimization procedure -- differential evolution (DE) -- which allows us to find periodic excitations resulting in a satisfactory level of resonance suppression.
The use of high-level algorithms (sometimes called metaheuristics) to optimize the final state of a time-dependent system over a range of energies has gained some traction in the quantum electrodynamics (QED) community in recent years, wherein the time-dependent Dirac equation is used to compute the time evolution of the system \cite{Kohlfurst2013, Hebenstreit2014, Hebenstreit2016}.
The approach used in this Letter is a combination of DE and time-independent Floquet calculations for periodic driving fields.
After giving a theoretical background of the problem, we show the effect of using non-monochromatic excitations on the Floquet absorption spectrum of a TLS.
Then, using the Fourier coefficients of the driving field as decision variables of a minimization problem, we find that it is possible to tailor this non-monochromatic field to suppress a given multiphoton resonance over a relatively large energy range.
Optimized solutions are characterized using their Floquet band structure and the effect of the optimization on the energy integrated spectrum is discussed.
The main finding of the Letter is that suppression occurs because of a dynamical gap closing (similar to CDT) which is the result of the coherent superposition of several Fourier harmonics and not of individual harmonics composing the driving function.
The application of this result to the fields of condensed-matter physics and quantum computing is discussed in conclusion.

Consider a TLS driven by a strong semi-classical field such that only the level populations are treated quantum mechanically.
Two-level dynamics are governed by the following $(2 \times 2)$ Hamiltonian \cite{Shevchenko20101, Jooya2016}
\begin{equation}
\label{eq:graphene2}
H(t) =  - \left[ \frac{\varepsilon_0}{2} + \frac{A}{2} g(t) \right] \sigma_z- \frac{\Delta}{2} \sigma_x.
\end{equation}
The first term can be interpreted as a time-dependent energy bias driving the quantum system, whereas the second term is the energy gap, or ``tunneling amplitude'' of the system \cite{Grifoni1998, Shevchenko20101}.
Depending on the exact nature of the quantum system, the values of the energy scales $\varepsilon_0$ and $\Delta$ may be related to dc and ac flux biases in a qubit \cite{Son2009, Jooya2016} or quasi-particle momenta in condensed-matter systems \cite{Fillion-Gourdeau2016, Gagnon2017}, among others.
The driving term $Ag(t)/2$ usually corresponds to an oscillating electric field, with $A$ being a normalized field amplitude related to the frequency of population oscillations (or Rabi frequency).
In this work, it is assumed that $H(t)$ and $g(t)$ are periodic with $T = 2 \pi / \omega$, where $\omega$ is the excitation frequency of the fundamental mode.
Thus, the following Fourier expansions hold:
\begin{equation}
\label{eq:gfourier}
g(t) = \sum_{n=-\infty}^{\infty} c_n e^{-in \omega t}, \quad H(t) = \sum_{n=-\infty}^{\infty} H^{[n]} e^{-in \omega t}.
\end{equation}
Substituting Eq. \eqref{eq:gfourier} in Eq. \eqref{eq:graphene2} yields the Fourier coefficients of the Hamiltonian
\begin{subequations}
	\begin{equation}
	H^{[0]} = - \left[ \frac{\varepsilon_0}{2} + \frac{A}{2} c_0 \right] \sigma_z - \frac{\Delta}{2} \sigma_x,
	\end{equation}
	\begin{equation}
	H^{[n]} = - \frac{A}{2} c_n \sigma_z, \qquad (n \neq 0).
	\end{equation}
\end{subequations}
For the purposes of this article, we consider odd periodic excitations (see supplementary material) which are represented by Fourier series of the form $g(t) = \sum_{n=1}^{\infty} b_n \sin n \omega t.$
This series can be recast into Eq. \eqref{eq:gfourier} using the substitution $c_n = - ib_n/2$ for $n > 0$ and $c_n = i b_{-n} /2$ for $n<0$.

Following the approach described in \cite{Son2009, Jooya2016} for superconducting flux qubits and extended to graphene in \cite{Gagnon2017}, one can introduce the Floquet state nomenclature in order to compute photo-excitation probabilities.
Floquet states are defined as $\ket{\nu n} = \ket{\nu} \otimes \ket{n}$,
where $\nu$ is the system index which can take two values, $\alpha$ and $\beta$, labeling negative and positive energy eigenstates of $\sigma_z$, respectively.
Switching to Fourier space and applying the Floquet theorem yields
\begin{equation}\label{eq:eigenvalue}
\sum_{\mu = \alpha, \beta} \sum_{m} \bra{\nu n} H_F \ket{\mu m} \left\langle \mu m | q_{\gamma l} \right\rangle = q_{\gamma l} \left\langle \nu n | q_{\gamma l} \right\rangle,
\end{equation}
where $q_{\gamma l}$ are the Floquet quasi-energies, $\ket{q_{\gamma l}}$ are the Floquet eigenvectors, and $H_F$ is the Floquet Hamiltonian.
The observable used in this article is the transition probability between field-free eigenstates $\ket{-}$ and $\ket{+}$, which have negative and positive eigenenergies, respectively (detailed definition is in Ref. \cite{Gagnon2017}).
This probability can be written as a sum of $k$-photon transition probabilities \cite{Son2009}
\begin{equation}
\label{eq:probability}
\bar{P}_{\ket{-} \rightarrow \ket{+}} = \sum_k \sum_{\gamma l} \left| \left\langle +, k | q_{\gamma l} \right\rangle \left\langle q_{\gamma l} | -, 0 \right\rangle \right|^2.
\end{equation}
It should be noted that, despite the apparent simplicity of the Floquet treatment, analytical solutions to Eq. \eqref{eq:eigenvalue} for a general driving function $g(t)$ do not exist \cite{Grifoni1998, Shevchenko20101}.
Therefore, transition probabilities must be evaluated numerically by computing the eigenvalues of a truncated version of the Floquet Hamiltonian \cite{Son2009, Gagnon2017}. 

\begin{figure}
	\includegraphics[width=0.5\textwidth]{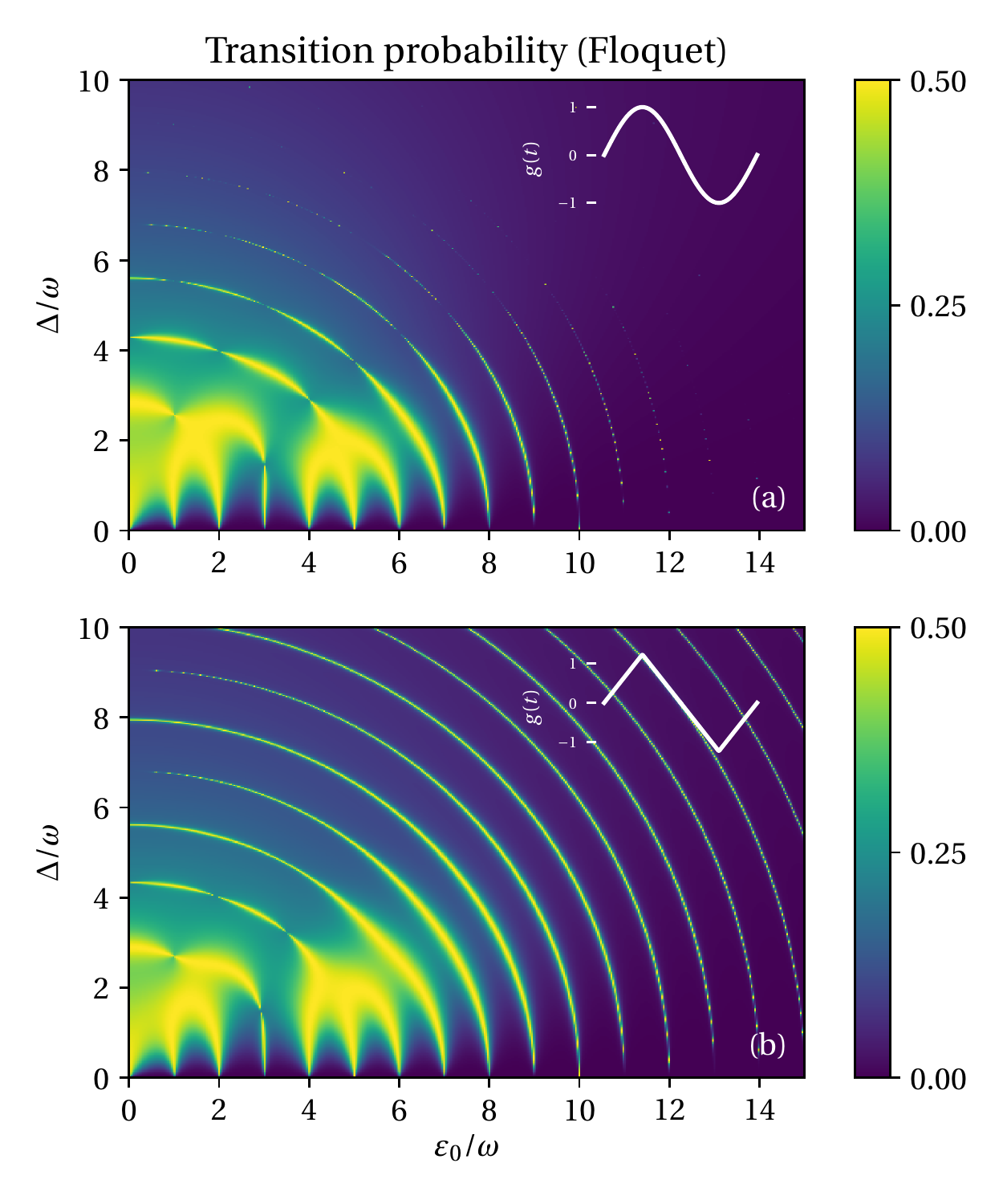}
	\caption{Floquet transition probability in a strongly driven TLS for (a) a monochromatic excitation and (b) a triangle wave. The oscillation frequency is set to $\omega$ with $A/\omega = 6$, and the normalization of $g(t)$ in the case of the triangle wave is chosen such that the amplitude of the fundamental harmonic is equal to 1.}
	\label{fig:momentum_maps}
\end{figure}

The spectral content of the periodic driving field has a manifest impact on the presence of multiphoton absorption peaks, as can be seen by comparing the Floquet transition probability resulting from a monochromatic excitation with that resulting from a non-monochromatic one (Fig. \ref{fig:momentum_maps}).
The transition probability induced by a monochromatic periodic excitation of frequency $\omega$ ($b_1 = 1$, $b_n =0$ for $n \neq 1$) is shown in Fig. \ref{fig:momentum_maps}a for a range of values of $\varepsilon_0$ and $\Delta$.
In the limit $\Delta^2 \ll \varepsilon_0^2, |A \omega|$, the field-free eigenstates can be approximated by those of $\sigma_z$, that is $\ket{\alpha}$ and $\ket{\beta}$ (diabatic basis).
Starting from Eq. \eqref{eq:probability}, a leading order perturbation treatment leads to the following analytic formula for the transition probability \cite{Oliver2005, Son2009}
\begin{equation}
\label{eq:probability2}
\bar{P}_{\ket{\alpha} \rightarrow \ket{\beta}} = \sum_k \frac{1}{2}\frac{[\Delta J_k(A /\omega)]^2}{[\Delta  J_k(A /\omega)]^2 + [k \omega - \varepsilon_0]^2 },
\end{equation}
where $J_k$ is the Bessel function of the first kind.
In short, for a monochromatic excitation the Floquet transition probability can be expressed as the superposition of Lorentzian $k$-photon resonances with an effective tunnel splitting given by $\tilde{\Delta} \equiv \Delta J_k(A /\omega)$ \cite{Grifoni1998}.
Values of $A /\omega$ corresponding to the $n$th zero of the Bessel function $J_k$, $j_{k,n}$, imply a vanishing effective splitting, leading to coherent destruction of tunneling (CDT) \cite{Grifoni1998, Son2009, Jooya2016, Gagnon2017}.
When the CDT condition is nearly satisfied, the Floquet transition probability exhibits narrow multiphoton resonances, as is the case for the $k=3$ peak in Fig. \ref{fig:momentum_maps}.

From a physical point of view, adding higher order harmonics to the driving signal corresponds to higher photon energies, which means high order multiphoton resonances can be efficiently excited.
This can be observed from the result displayed in Fig. \ref{fig:momentum_maps}b, where the function $g(t)$ is a triangle wave with a fundamental mode amplitude equal to 1, i.e. $b_n = \frac{(-1)}{n^2}^{(n-1)/2}$ for $n=1,3,5 \ldots$.
Although high order multiphoton resonances are brighter in the case of triangle wave driving, the structure of the low-order rings is mostly similar to the case of the monochromatic excitation, including the resonances' width (see Fig. \ref{fig:momentum_maps}a).

Now that the effect of two exemplary driving functions on multiphoton absorption has been established, we turn our attention to the inverse problem, namely finding a driving function that results in a chosen multiphoton absorption feature.
To achieve this, we seek values of the Fourier coefficients in Eq. \eqref{eq:gfourier} which result in suppression of the Floquet transition probability over some pre-defined range $\mathcal{D}$ of values of $\varepsilon_0$ and $\Delta$.
This can be written in mathematical terms as a minimization problem
\begin{equation}
F = \min_{\vec{X} \in \mathbb{R}^N} f(\vec{X})
\end{equation}
where $F$ is a minimum of the objective function $f$ in parameter space, $\vec{X} = \big\lbrace X_1, X_2 , \cdots , X_N \big\rbrace$ is a decision vector composed of $N$ optimization variables, and $f(\vec{X})$ is chosen as the cumulative Floquet transition probability:
\begin{equation}
f = \iint_{\mathcal{D}}  \bar{P}_{\ket{-} \rightarrow \ket{+}} ~ d \varepsilon_0 d \Delta.
\end{equation}
Appropriate upper and lower bounds are imposed on the values of the optimization variables (see supplementary material), and those are related to the Fourier coefficients in Eq. \eqref{eq:gfourier} via $b_{2n - 1} = X_n$,
with even Fourier coefficients set to 0 by choice.
We seek solutions with only odd order Fourier harmonics and alternating signs in order to obtain excitations relatively close to a sine wave or a triangle wave.
This preserves the stability of the numerical method (see supplementary material for more details).

In this Letter, it is assumed that the effect of the individual decision variables (individual harmonics) on the Floquet transition probability cannot be isolated.
Consequently, the search space of the optimization problem is a $N-$dimensional hypercube, where $N$ is the number of decision variables.
Since searching this hypercube requires multiple objective function evaluations which in turn imply a numerical computation of the transition probability, we seek local minima of the optimization problem using differential evolution (DE), a multi-purpose optimization algorithm \cite{Talbi2009}.
Specifically we use a parallel version of DE \cite{Storn1997,das2011differential} implemented in the \textsc{Pagmo} optimization library \cite{izzo2012pygmo}.
In a nutshell, DE is a population based, evolutionary optimization algorithm based on mutation and recombination operators which direct the search towards good solutions using vector differences \cite{Talbi2009}.
The \textsc{Pagmo} library uses the Generalized Island model (GIM) for parallelization \cite{Rucinski2010555,Izzo2012}.
According to this model, optimization takes place on different ``islands'' (corresponding to an execution thread) in parallel, and on each of these islands, an instance of DE is executed.
Information on optimal solutions is transferred between the islands after each generation (see supplementary material).

\begin{figure}
	\includegraphics[width=0.5\textwidth]{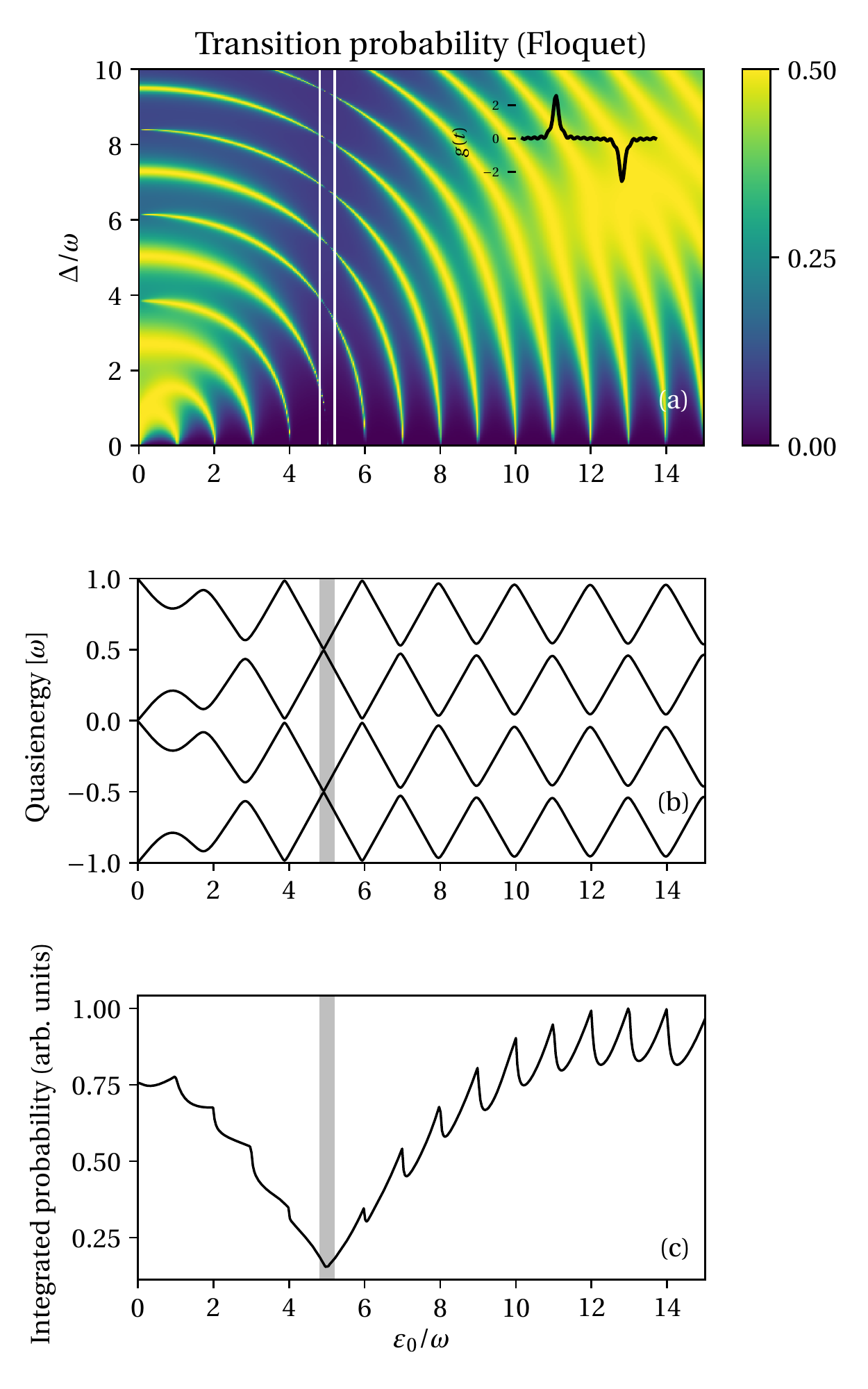}
	\caption{Suppression of the $k=5$ multiphoton resonance using pulse shape optimization with $A/\omega = 9$ and 10 optimization variables (Fourier coefficients).
		(a) Floquet transition probability plot showing the range over which suppression takes place (white rectangle) as well as the corresponding optimized driving function (inset).
		(b) Evolution of the Floquet quasi-energies for $\Delta / \omega = 4$.
		The shaded rectangle indicates the region over which the transition probability is minimized, which corresponds to a close encounter of quasi-states.
		(c) Transition probability integrated over $\Delta / \omega$, showing the suppressed $k=5$ multiphoton peak.}
	\label{fig:suppression}	
\end{figure}

Using DE, it is possible to find the Fourier coefficients of an excitation which suppresses a given multiphoton resonance in the Floquet absorption spectrum: this result is shown in Fig. \ref{fig:suppression}.
The optimization target is the minimization of the cumulative Floquet transition probability over the range $\mathcal{D} = \{4.8 \leq \varepsilon_0/\omega \leq 5.2, \, 0 \leq \Delta/\omega \leq 10 \}$, indicated by a white rectangle on Fig. \ref{fig:suppression}.
This corresponds to the suppression of the 5-photon absorption peak, which is selected because of its broadness in the case of a monochromatic or triangle excitation (see Fig. \ref{fig:momentum_maps}).
In this optimization run, we used $A/\omega = 9$ with the following bounds for the fundamental harmonic: $b_1 \in [0.66, 1.0]$.
After setting appropriate upper and lower bounds on the other coefficients (see supplementary material), the value of the decision vector $\vec{X}$ is optimized using parallel DE. 
We set the number of decision variables to 10 and use 48 parallel islands each containing a population size of 10 individuals (see supplementary material for details).
The optimization routine evolves populations for 20 generations on each island.
The value of $g(t)$ which results in the best local minimum is shown in the inset of Fig. \ref{fig:suppression}a, which also shows the corresponding multiphoton fringes.
The $k=5$ peak is the most suppressed resonance.

The suppression of the transition probability can be directly related to a closing of the dynamical gap between Floquet eigenstates, as can be seen in Fig. \ref{fig:suppression}b, which shows the Floquet band diagrams (quasi-energies as a function of $\varepsilon_0 / \omega$) for a fixed value of $\Delta / \omega = 4$.
We have verified that the dynamical gap closing over the whole energy range is the result of the coherent superposition of several Fourier harmonics, and not of individual harmonics composing the signal.
This was achieved by computing the transition probability for driving functions containing only the fundamental ($b_1 = 0.66$) or the second ($b_3 = -0.518$, as found by the DE algorithm) harmonic, and we found that the $k=5$ resonance is not suppressed in these cases (not shown).
This can also be checked using the analytical result, Eq. \eqref{eq:probability2}, for $\Delta / \omega \rightarrow 0$.
Moreover, we have verified that considering a monochromatic excitation with the same peak value as the optimized pulse (displayed in the inset of Fig. \ref{fig:suppression}a) does not result in a suppression of the resonance.
In short, the coherent superposition of several Fourier harmonics results in CDT without the fundamental mode amplitude itself (or other harmonics) satisfying the CDT condition described earlier for the $k=5$ resonance.
This is confirmed by the fact that increasing the number of decision variables results in progressively stronger suppression (see supplementary material for details).
Conversely, the $k=3$ multiphoton resonance, which was very narrow in the case of a purely monochromatic excitation for a value of $A/\omega = 6$ (see Fig. \ref{fig:momentum_maps}a), is much broader in the case of the optimized excitation (which corresponds to an effective field strength of $A/\omega = 5.94$ if one only considers the fundamental harmonic).

The effect of the transition probability suppression can be seen clearly in the energy-integrated spectrum, shown in Fig. \ref{fig:suppression}c.
This result shows the cumulative transition probability integrated over all values of $\Delta$, and shows a dip where the $k=5$ multiphoton peak should appear.
This spectrum could correspond for instance to integration over transverse quasi-particle momenta in graphene, which could be obtained experimentally using photoemission spectroscopy \cite{Freericks2009, Ishikawa2013}.
As described in Ref. \cite{Gagnon2017}, integrating the Floquet transition probability $\bar{P}_{\ket{\alpha} \rightarrow \ket{\beta}}$ over the whole momentum space (all values of $\varepsilon_0$ and $\Delta$) corresponds to photo-induced carrier densities of the order $\bar{n} \simeq 10^{11}$ cm$^{-2}$, if one considers laser frequencies in the THz range \cite{Gagnon2017}.

The results of this Letter could find application in the control of scattering mechanisms in condensed-matter physics, since scattering mechanisms have various functional dependences on the excess carrier density $\bar{n}$ in Dirac materials such as graphene \cite{DasSarma2011}.
At low temperatures, the main scattering channel in monolayer graphene at high carrier densities is short-range disorder.
In contrast, as the carrier density $\bar{n}$ increases, Coulomb interactions become screened and thus charged impurity scattering decreases in importance.
The effect of both these scattering channels has been experimentally quantified for chemical-vapor-deposited (CVD) graphene in a recent article by Yu \emph{et al.} \cite{Yu2016}.
Pulse shape optimization paves the way to light-matter experiments with mono-layer graphene wherein short-range disorder or charged impurity scattering can be minimized over a momentum window, allowing one to measure observables other than the pair production probability.

The theoretical findings of this Letter could also be applied to the field of quantum computing.
Indeed, Deng \emph{et al.} have recently proposed the suppression of transitions between Floquet states as the basis for a new control protocol which enables the design of high-fidelity quantum gates \cite{Deng2016}.
An attractive feature of our central result is that the Floquet transition probability is suppressed over a relatively large energy range, potentially allowing for a wider range of operation conditions of the quantum gate.

In summary, we find that it is possible to selectively suppress multiphoton absorption peaks in driven two-level systems using spectral optimization.
Our main finding, obtained using a combination of a metaheuristic optimization algorithm and Floquet theory, is that resonance suppression is due to a coherent superposition of several Fourier harmonics which do not individually satisfy the conditions of coherent destruction of tunneling.
Although the results obtained from Floquet theory are restricted to periodic excitations in the narrow sense, they might constitute a starting point for optimization using time-dependent calculations.
In actuality, the momentum space patterns obtained from the two approaches agree well for a pulse of finite duration, e.g. 10 cycles \cite{Fillion-Gourdeau2016}.
This work may also stimulate the comparison of Floquet methods to solutions of geometric and numerical ensemble control in cases where periodic solutions are satisfactory.

\begin{acknowledgments}
	DG is supported by a  postdoctoral research scholarship from Fonds de recherche du Qu\'ebec -- Nature et technologies (FRQNT).
	Computations were made on the supercomputer \emph{Mammouth} from Universit\'e de Sherbrooke, managed by Calcul Qu\'ebec and Compute Canada.
	The operation of this supercomputer is funded by the Canada Foundation for Innovation (CFI), minist\`ere de l'\'{E}conomie, de la Science et de l'Innovation du Qu\'{e}bec (MESI) and FRQNT.
\end{acknowledgments}

\bibliography{cond-mat}
%\bibliography{extracted}

\end{document}